# Dirac's Representation Theory as a Framework for Signal Theory. II. Infinite Duration and Continuous Signals

Alexander Gersten

Department of Physics,

and Unit of Biomedical Engineering, and Zlotowski Center for Neuroscience,

Ben-Gurion University of the Negev

Beer-Sheva 84105, Israel

e-mail: gersten@bgumail.bgu.ac.il

## Abstract

In the preceding paper [1] we dealt with discrete signals of finite duration. Here we generalize previous results and demonstrate that the Dirac representation theory can be effectively adjusted and applied to continuous or discrete signals of infinite time duration. The role of the identity and projection operators is emphasized. The sampling theorem is viewed from the point of view of orthogonal physical states. An orthogonal basis which spanned the time space, ceases to be orthogonal and becomes overcomplete if the domain of frequencies is restricted in a bandwidth. In this case there exists an infinite number of sub-bases of discrete times which are orthogonal and complete. The relation between the overcomplete bases and a complete one is the essence of the sampling theorem. The signal theory is reformulated in the framework of the Dirac bra-kets. The case of signals existing for positive time is treated in detail.

## 1. Introduction

The usual approach to signal theory is to start with the continuous case and treat the discrete finite one as an approximation. In the preceding paper [1], which we will denote by [I], we have adopted a different approach for the following reasons:

1. In practice the data are collected in finite discrete sequences [2].
2. The discrete finite case has its own peculiarities, different from the continuous case. Moreover, there are several, possibly many, discrete cases for each continuous case. The continuous case, if it exists, is a common limit of different discrete cases [3-5].
3. The transition to the continuos case is more complicated, as the limits can be distributions (generalized functions) [6-8].

Most of the developments in this paper are based on the assumption of Dirac [9], made in 1930, that a given self-adjoint operator $\hat{A}$ can be presented as

$$\hat{A} = \int_{\sigma(\hat{A})} \alpha |\alpha\rangle\langle\alpha| d\alpha, \qquad (1.1)$$

where $\hat{A}|\alpha\rangle = \alpha|\alpha\rangle$, and $\sigma(\hat{A})$ is the spectrum of $\hat{A}$. Dirac assumed that all properties of hermitean operators on finite dimensional vector spaces can be extended and be valid in infinite dimensional Hilbert spaces even for unbounded self-adjoint operators. A full mathematical justification of the Dirac formalism came only about thirty years later. At the beginning the formalism was severely criticized by mathematicians.

Von Neumann in the preface to his 1932 fundamental book [3]: "Mathematical Foundations of Quantum Mechanics" wrote: "Dirac, in several papers, as well as in his recently published book, [2] has given a representation of quantum mechanics which is scarcely to be surpassed in brevity and elegance, and which is at the same time of invariant character. It is therefore perhaps fitting to advance a few arguments on behalf of our method, which deviates considerably from that of Dirac.

The method of Dirac, mentioned above, (and this is overlooked today in a great part of quantum mechanical literature, because of the clarity and elegance of the theory) in no way satisfies the requirements of mathematical rigor -- not even if these are reduced in a natural and proper fashion to the extent common elsewhere in theoretical physics. For example, the method adheres to the fiction that each self-adjoint operator can be put in diagonal form. In the case of those operators for which this is not actually the case, this requires the introduction of "improper" functions with self-contradictory properties. The insertion of such a mathematical "fiction" is frequently necessary in Dirac's approach, even though the problem at hand is merely one of calculating numerically the result of a clearly defined experiment. There would be no objection here if these concepts, which cannot be incorporated into the present day framework of analysis, were intrinsically necessary for the physical theory. Thus, as Newtonian mechanics first brought about the development of the infinitesimal calculus, which, in its original form, was undoubtedly not self consistent, so quantum mechanics might suggest a new structure for our "analysis of infinitely many variables" -- i.e., the mathematical technique would have to be changed, and not the physical theory. But this is by no means the case. It should rather be pointed out that the quantum mechanical 'Transformation theory" can be established in a manner which is just as clear and unified, but which is also without mathematical objections. It should be emphasized that the correct structure need not consist in a mathematical refinement and explanation of the Dirac method, but rather that it requires a procedure differing from the very beginning, namely, the reliance on the Hilbert theory of operators."

The above citation was written before the invention of distributions [1c],[1d] in the 1950's. The theory of distributions gave new insight into the Dirac formalism and relieved it from the accusation of using improper functions. A complete justification of the Dirac formalism was given by Gel'fand [10-12], who introduced the rigged Hilbert space and the generalized eigenvectors [10-13]. Within the framework of this theory, eq. (1.1) is fully justified and can be applied in a simple manner, which generalises the finite-dimensional results of paper [I]. The theory of Gel'fand is also the basis of axiomatic quantum field theory [14] and was further adapted and elaborated in order to describe more plainly quantum mechanics in the Dirac formalism [15-20].

In the treatment of the finite discrete signals we used the bra-ket formalism [9] and ket bases to describe the transition from signals represented by time points to their frequency representation. As a guideline we used the orthonormality condition, which is the condition that physical states (eigenstates of self-adjoint operators) satisfy in quantum mechanics (in one dimension, and for different eigenvalues). The peculiarity of using ket-vectors and not wavefunctions is that orthogonality can be employed (at least for the problems of this paper), without the need to explicitly refer to boundary conditions. One of the reasons is that wavefunctions in Hilbert space are required to vanish at space infinity, while for ket-vectors (in the rigged Hilbert space) space infinity is a regular point. We continue here the same approach and consider, for example, the sampling theorem from this point of view, and we find that the use of orthogonality and projection operators lead us to understand the sampling theorem as coming from a relation between an overcomplete non-orthogonal basis and an orthonormal complete one. Other aspects of signal theory are treated as well in the bra-ket formalism.

In this paper we introduce the Dirac representation as a framework or a basis for signal theory. As the subject is quite broad we will concentrate in the paper only on theoretical considerations barely touching applications, leaving them for next publications.

## 2 The infinite dimensional Dirac space

In paper [I] we considered a basis with a discrete and finite index. The treatment of the basis with a continuous index is more involved. For the continuous case the ket basis (I3.4) can be generalized to a ket basis labeled by the time t :

$$|t\rangle, \text{ where: } -\infty \leq t \leq \infty, \qquad (2.1a)$$

with the dual bra space:

$$\langle t|, \text{ where: } -\infty \leq t \leq \infty, \qquad (2.1b)$$

for which there exists the scalar product (in terms of distributions [6-8]):

$$\langle t|t'\rangle = \delta(t-t'), \qquad (2.2)$$

where $\delta$ is the Dirac delta function (see Appendix A). The identity operator now takes the following form:

$$\hat{I} = \int_{-\infty}^{\infty} |t\rangle\langle t|dt. \tag{2.3}$$

Equations (2.1-3) define the largest Dirac space for continuous signals. In a way similar to eq. (I-2.1), a ket-vector and a bra-vector in this space (a rigged Hilbert space ) will have the most general expansion with the bases (2.1a) and (2.1b) in the form:

$$|u\rangle = \int_{-\infty}^{\infty} u(t)|t\rangle dt, \quad \langle u| = \int_{-\infty}^{\infty} (u(t))^* \langle t|dt, \tag{2.4}$$

respectively, where u(t) are the expansion coefficients, which can be distributions. From the orthonormality condition (2.3) one gets:

$$\langle t|u\rangle = u(t), \quad \langle u|t\rangle = (u(t))^*. \tag{2.5}$$

Let us take as an example the signal f(t), which can be considered as a ket $|f\rangle$ with projections into the orthonormal basis $|t\rangle$ (i.e. the components) equal to the values of the function, i.e.:

$$\langle t|f\rangle = f(t). \tag{2.6}$$

In the Dirac space one can define other orthonormal ket-bases, among them the one labeled with the angular frequency ω: $|\omega\rangle$. Its transition to the t-basis is obtained via the matrix elements similar to eq. (I-3.8):

$$\langle t|\omega\rangle = \frac{1}{\sqrt{2\pi}} e^{i\omega t}, \tag{2.7}$$

from which we can get that

$$|\omega\rangle = \frac{1}{\sqrt{2\pi}} \int_{-\infty}^{\infty} e^{i\omega t}|t\rangle dt. \tag{2.8}$$

The normalization in eq. (2.7) was so chosen to ensure the orthonormality condition

$$\langle \omega|\omega'\rangle = \frac{1}{2\pi} \int_{-\infty}^{\infty} e^{i(\omega-\omega')t} dt = \delta(\omega-\omega'). \tag{2.9}$$

As the ω-basis spans the same Dirac space as the t-basis, the identity operator satisfies:

$$\hat{I} = \int_{-\infty}^{\infty} |t\rangle\langle t|dt = \int_{-\infty}^{\infty} |\omega\rangle\langle \omega|d\omega. \tag{2.10}$$

Similar to eq. (I-4.6) we can introduce the angular frequency operator:

$$\hat{\omega} = \int\limits_{-\infty}^{\infty} \omega |\omega\rangle\langle\omega| d\omega, \tag{2.11}$$

having the following properties

$$\hat{\omega}|\omega\rangle = \omega|\omega\rangle, \quad \langle\omega|\hat{\omega} = \langle\omega|\omega, \quad \hat{\omega}^n|\omega\rangle = \omega^n|\omega\rangle. \tag{2.12}$$

Using eqs. (2.10) and (2.7-9) one can prove that

$$\langle t|(i\hat{\omega})|u\rangle = \int\limits_{-\infty}^{\infty} \langle t|(i\hat{\omega})|\omega\rangle\langle\omega|u\rangle d\omega = \int\limits_{-\infty}^{\infty} i\omega\langle t|\omega\rangle\langle\omega|u\rangle d\omega$$
$$= \frac{1}{\sqrt{2\pi}} \int\limits_{-\infty}^{\infty} i\omega e^{i\omega t}\langle\omega|u\rangle d\omega = \frac{d}{dt} \int\limits_{-\infty}^{\infty} \langle t|\omega\rangle\langle\omega|u\rangle d\omega = \frac{d}{dt}\langle t|u\rangle, \tag{2.13}$$

i.e. in that sense the $i\hat{\omega}$ operator can be considered as the time derivative operator.

Let us introduce the time operator:

$$\hat{t} = \int\limits_{-\infty}^{\infty} t|t\rangle\langle t| dt, \tag{2.14}$$

having the following properties

$$\hat{t}|t\rangle = t|t\rangle, \quad \langle t|\hat{t} = \langle t|t, \quad \hat{t}^n|t\rangle = t^n|t\rangle. \tag{2.15}$$

In a way similar to the derivation of eq. (2.13) we can get:

$$\langle\omega|(-i\hat{t})|u\rangle = \frac{d}{d\omega}\langle\omega|u\rangle. \tag{2.16}$$

From eq. (2.7) we obtain

$$-i\frac{d}{dt}\langle t|\omega\rangle = \omega\langle t|\omega\rangle, \tag{2.17}$$

and more generally (rewriting eq. (2.13)):

$$-i\frac{d}{dt}\langle t|u\rangle = \langle t|\hat{\omega}|u\rangle, \tag{2.18}$$

i.e. according to eqs. (2.17) and (2.18) one can consider the operator

$$-i\frac{d}{dt} \equiv \breve{\omega}, \tag{2.19}$$

also as the angular frequency operator. The difference between the operators of eq. (2.11) and eq. (2.19) is that the operator $\hat{\omega}$ is an infinite dimensional matrix in the Dirac space, while the operator $\breve{\omega}$ acts in the space of functions. In the following we will deal mostly with operators in the Dirac space.

## 3. Continuous Spectral Analysis

The spectral decomposition of a time dependent signal f(t) is given by its Fourier transform:

$$F(\omega) = \frac{1}{\sqrt{2\pi}} \int_{-\infty}^{+\infty} e^{-it\omega} f(t) dt ,  \qquad (3.1)$$

while the inverse relation is:

$$f(t) = \frac{1}{\sqrt{2\pi}} \int_{-\infty}^{+\infty} e^{it\omega} F(\omega) d\omega .  \qquad (3.2)$$

Here we will utilize the Dirac bra-ket formalism [9], (again, in the rigged Hilbert space [18]), for spectral analysis. In practice the data are collected in finite samplings. In this section, for methodological reasons, we will confine ourselves to continuous signals.

We will denote the signal function f(t) as a bra-ket $\langle t | f \rangle$. Thus the signal is considered to be an abstract ket-vector state $|f\rangle$ and can be represented either by $\langle t | f \rangle$ or by its Fourier transform $\langle \omega | f \rangle$ :

$$\langle \omega | f \rangle = \int_{-\infty}^{+\infty} \langle \omega | t \rangle \langle t | f \rangle dt , \qquad (3.3)$$

and the inverse transform is

$$\langle t | f \rangle = \int_{-\infty}^{+\infty} \langle t | \omega \rangle \langle \omega | f \rangle d\omega , \qquad (3.4)$$

where:

$$F(\omega) \equiv \langle \omega | f \rangle , \qquad (3.5)$$

$$\langle t | \omega \rangle = \frac{1}{\sqrt{2\pi}} e^{it\omega} , \quad \langle \omega | t \rangle = \frac{1}{\sqrt{2\pi}} e^{-it\omega} , \qquad (3.6)$$

and the normalization condition is:

$$\langle t | t' \rangle = \int_{-\infty}^{\infty} \langle t | \omega \rangle \langle \omega | t' \rangle d\omega = \frac{1}{2\pi} \int_{-\infty}^{\infty} e^{i\omega(t-t')} d\omega = \delta(t-t') , \qquad (3.7)$$

where is the delta function of Dirac. In the Dirac formalism the completeness relation is:

$$\hat{I} = \int\limits_{-\infty}^{+\infty} |t\rangle\langle t|dt = \int\limits_{-\infty}^{+\infty} |\omega\rangle\langle\omega|d\omega \tag{3.8}$$

where $\hat{I}$ is the identity operator. The relations (3.6) are solutions of the eigenvalue equation (2.17)

### 3a. Windowing

In practice the signal is detected during a finite time interval, say 2T. The spectral analysis in this time interval can be done in the following way:

$$F(\omega)_T \equiv \langle\omega|\hat{A}|f\rangle_T = \tfrac{1}{\sqrt{2\pi}} \int\limits_{-T}^{T} \alpha(t)\exp(-it\omega)f(t)dt \tag{3.9}$$

where $\alpha(t)$ is one of possible windowing functions [22-25], [29] and eq. (3.9) was obtained by using the filter operator

$$\hat{A} = \int\limits_{-T}^{T} \alpha(t)|t\rangle dt\langle t|. \tag{3.9a}$$

In this case the formalism can be replaced by decomposing the identity operator into:

$$\hat{I} = \hat{A} + (\hat{I} - \hat{A}) = \int\limits_{-\infty}^{+\infty} \alpha(t)|t\rangle\langle t|dt + \int\limits_{-\infty}^{+\infty} [1-\alpha(t)]|t\rangle\langle t|dt, \tag{3.8a}$$

where $\alpha(t)$ is the windowing function having support only in the finite interval. For example for the rectangular window with a finite time interval 2T:

$$\alpha(t) = 1, \quad \text{for} \quad -T \leq t \leq T; \quad (\text{otherwise} \quad \alpha(t) = 0.) \tag{3.10}$$

One may employ also other windowing procedures of common use. Substituting eq. (3.8a) into eq(3.3) we obtain:

$$\langle\omega|f\rangle = \int\limits_{-\infty}^{+\infty} \alpha(t)\langle\omega|t\rangle\langle t|f\rangle dt + \int\limits_{-\infty}^{+\infty} [1-\alpha(t)]\langle\omega|t\rangle\langle t|f\rangle dt$$

$$= \int\limits_{-\infty}^{+\infty} \alpha(t)\langle\omega|t\rangle\langle t|f\rangle dt + \int\limits_{-\infty}^{+\infty} d\omega' \int\limits_{-\infty}^{+\infty} [1-\alpha(t)]\langle\omega|t\rangle\langle t|\omega'\rangle\langle\omega'|f\rangle dt$$

$$= \langle\omega|f\rangle_T + \int\limits_{-\infty}^{+\infty} K(\omega,\omega')\langle\omega'|f\rangle d\omega', \tag{3.11}$$

where

$$K(\omega,\omega') = \int\limits_{-\infty}^{+\infty} [1-\alpha(t)]\langle\omega|t\rangle\langle t|\omega'\rangle dt \qquad (3.12)$$

is the kernel of the linear integral equation (3.11), and

$$\langle\omega|f\rangle_T = \int\limits_{-\infty}^{+\infty} \alpha(t)\langle\omega|t\rangle\langle t|f\rangle dt \qquad (3.13)$$

is the spectrum of the windowed signal (the measured quantity). Equation (3.11) can be recast in an operatorial form if we substitute: $K(\omega,\omega') = \langle\omega|\hat{K}|\omega'\rangle \equiv \langle\omega|\hat{I} - \hat{A}|\omega'\rangle$, and thus obtain from eq. (3.11):

$$\langle\omega|f\rangle = \langle\omega|f\rangle_T + \int\limits_{-\infty}^{+\infty} \langle\omega|\hat{K}|\omega'\rangle\langle\omega'|f\rangle d\omega'. \qquad (3.14)$$

Eq.(3.14) can be used for an approximate estimation of the error of the spectral analysis in the following way:

$$\langle\omega|f\rangle - \langle\omega|f\rangle_T \approx \int \langle\omega|\hat{K}|\omega'\rangle\langle\omega'|f\rangle_T d\omega', \qquad (3.15)$$

and when possible (i.e. if eq. (3.14) has a solution), to get the exact spectrum form the windowed one. In the Appendix B we give an example of a continuous signal with windowing.

### 3b. Filtering

The filtering can be described with the help of the filter operator

$$\hat{\Phi} = \int\limits_{-\infty}^{\infty} \varphi(\omega)|\omega\rangle d\omega\langle\omega|, \qquad (3.16a)$$

which is an operator valued, continuous, linear functional. The filtered spectrum is:

$$\langle\omega|\hat{\Phi}|f\rangle = \varphi(\omega)\langle\omega|F\rangle = \int\limits_{-\infty}^{\infty} \langle\omega|t\rangle\langle t|\hat{\Phi}|f\rangle dt = \int\limits_{-\infty}^{\infty} \langle\omega|t\rangle\langle t|f\rangle_\varphi dt, \qquad (3.16b)$$

where the function $\varphi(\omega)$ modifies (filters) the spectrum and $\langle t|f\rangle_\varphi = \langle t|\Phi|f\rangle$ is the filtered signal. In a similar way to that derived in sec. 3a, one can derive the obvious integral equation which relates the signal to the filtered signal:

$$\langle t|f\rangle = \langle t|\hat{\Phi}|f\rangle + \langle t|(\hat{I}-\hat{\Phi})|f\rangle = \langle t|\hat{\Phi}|f\rangle + \int\limits_{-\infty}^{\infty} \langle t|(\hat{I}-\hat{\Phi})|t'\rangle\langle t'|f\rangle dt'. \qquad (3.16c)$$

Let:

$$\varphi(\omega) = \int_{-\infty}^{\infty} \langle\omega|t\rangle\langle t|\varphi\rangle dt, \qquad (3.17)$$

then, using eq. (3.17), the inverse transform of (3.16) is

$$\langle t|f\rangle_\varphi = \int_{-\infty}^{\infty} \langle t|\omega\rangle\varphi(\omega)\langle\omega|f\rangle d\omega = \int_{-\infty}^{\infty}\int_{-\infty}^{\infty} \langle t|\omega\rangle\langle\omega|t'\rangle\langle t'|\varphi\rangle\langle\omega|f\rangle d\omega dt'. \qquad (3.18)$$

From eqs. (3.6) we have:

$$\langle t|\omega\rangle\langle\omega|t'\rangle = \langle t-t'|\omega\rangle,$$

and eq. (3.18) becomes, after using the r.h.s of eq. (3.8):

$$\langle t|f\rangle_\varphi = \int_{-\infty}^{\infty} \langle t'|\varphi\rangle\langle t-t'|f\rangle dt' = \int_{-\infty}^{\infty} \langle t-t'|\varphi\rangle\langle t'|f\rangle dt' \qquad (3.19)$$

in which we recognize the convolution theorem. The inverse of eq. (3.17) is

$$\langle t|\varphi\rangle = \int_{-\infty}^{\infty} \varphi(\omega)(1/\sqrt{2\pi})\exp(it\omega)d\omega. \qquad (3.20)$$

Let us consider for example:

$$\varphi(\omega) = \begin{cases} 1 & \text{for} \quad \omega_1 < \omega < \omega_2 \quad \text{and} \quad -\omega_2 < \omega < -\omega_1 \\ 0, & \text{otherwise,} \end{cases} \qquad (3.21)$$

than

$$\langle t|f\rangle_\varphi = \sqrt{\frac{2}{\pi}}\int_{-\infty}^{\infty} \frac{[\sin((t-t')\omega_2) - \sin((t-t')\omega_1)]\langle t'|f\rangle}{(t-t')} dt'. \qquad (3.22)$$

If the frequencies of the signal are expected to be in the above mentioned region, i.e. bounded by eq. (3.21), then $\langle t|f\rangle_\varphi \equiv \langle t|\hat{\Phi}|f\rangle = \langle t|f\rangle$ and eq. (3.22) becomes an identity

$$\langle t|\hat{\Phi}|f\rangle = \sqrt{\frac{2}{\pi}}\int_{-\infty}^{\infty} \frac{[\sin((t-t')\omega_2) - \sin((t-t')\omega_1)]\langle t'|\hat{\Phi}|f\rangle}{(t-t')} dt', \qquad (3.22a)$$

which is a particular case of eq. (A.9) in Appendix A. Later we will relate this result to the sampling theorem.

### 3c. Averaging

One can further extend the Dirac formalism and obtain easily both well known, and as well, less known results. For example:

$$\langle f|f\rangle = \int_{-\infty}^{\infty} \langle f|t\rangle\langle t|f\rangle dt = \int_{-\infty}^{\infty} \langle f|\omega\rangle\langle\omega|f\rangle d\omega, \qquad (3.23)$$

is obtained directly from eq. (3.8) and is the well known Parseval theorem. One can define average quantities in a manner similar to that done in quantum mechanics, e.g.:

$$\langle\hat{\omega}^n\rangle = \frac{\langle f|\hat{\omega}^n|f\rangle}{\langle f|f\rangle} = \frac{\int_{-\infty}^{\infty}\langle f|\hat{\omega}^n|\omega\rangle\langle\omega|f\rangle d\omega}{\langle f|f\rangle} = \frac{\int_{-\infty}^{\infty}\langle f|\omega\rangle\omega^n\langle\omega|f\rangle d\omega}{\langle f|f\rangle}, \qquad (3.24a)$$

or work directly with the signal f(t) using eq. (2.18):

$$\langle\hat{\omega}^n\rangle = \frac{\int_{-\infty}^{\infty}\langle f|t\rangle\langle t|\hat{\omega}^n|f\rangle dt}{\langle f|f\rangle} = \frac{\int_{-\infty}^{\infty}\langle f|t\rangle(-i\frac{d}{dt})^n\langle t|f\rangle dt}{\langle f|f\rangle}. \qquad (3.24b)$$

In order to define an average signal we introduce the signal operator:

$$\hat{f} = \int_{-\infty}^{\infty} f(t)|t\rangle\langle t|dt, \qquad \hat{f}|t\rangle = f(t)|t\rangle, \qquad (3.25)$$

with the above definition the average will be

$$\langle\hat{f}\rangle = \frac{\langle f|\hat{f}|f\rangle}{\langle f|f\rangle} = \frac{\int_{-\infty}^{\infty}\langle f|t\rangle\langle t|f\rangle\langle t|f\rangle dt}{\langle f|f\rangle}. \qquad (3.26)$$

An analog of mean kinetic energy will be

$$K = \left\langle\left[\frac{d}{dt}\langle t|f\rangle\right]^2\right\rangle = \frac{\int_{-\infty}^{\infty}\langle f|t\rangle\left[\frac{d}{dt}\langle t|f\rangle\right]^2\langle t|f\rangle dt}{\langle f|f\rangle}, \qquad (3.27)$$

We can see that the introduction of the Dirac formalism in Fourier and signal analyses can lead to simplifications and to new definitions of averages.

### 3d. The Autocorrelation and the Wigner Double Distribution Functions

Let us consider the square of the absolute value of the signal S(t):

$$|\langle t|S\rangle|^2 = \langle S|t\rangle\langle t|S\rangle = \int_{-\infty}^{\infty}\int_{-\infty}^{\infty} \langle S|\omega\rangle\langle\omega|t\rangle\langle t|\omega'\rangle\langle\omega'|S\rangle d\omega d\omega'$$
$$= \frac{1}{2\pi}\int_{-\infty}^{\infty}\int_{-\infty}^{\infty} \langle S|\omega\rangle\langle\omega'|S\rangle\exp(it(\omega'-\omega))d\omega d\omega' \qquad (3.31)$$

and substitute $\omega = \omega' + \tau$, $\omega' = \vartheta - \tau/2$, we obtain:

$$|\langle t|S\rangle|^2 = \frac{1}{2\pi}\int_{-\infty}^{\infty}\int_{-\infty}^{\infty}\langle S|\vartheta+\tau/2\rangle\langle\vartheta-\tau/2|S\rangle\exp(-it\tau))d\vartheta d\tau, \qquad (3.32)$$

eq. (3.32) introduces the point-splitting regularization which gives the Wigner function, which we derive next. Depending on the order of integration, we get two very important functions

$$|\langle t|S\rangle|^2 = \frac{1}{2\pi}\int_{-\infty}^{\infty}R(\tau)\exp(-it\tau))d\tau, \qquad (3.33)$$

$$|\langle t|S\rangle|^2 = \frac{1}{2\pi}\int_{-\infty}^{\infty}W_1(t,\theta)d\vartheta, \qquad (3.34)$$

where:

$$R(\tau) = \frac{1}{2\pi}\int_{-\infty}^{\infty}\langle S|\vartheta+\tau/2\rangle\langle\vartheta-\tau/2|S\rangle d\vartheta, \qquad (3.35)$$

$$W_1(t,\vartheta) = \frac{1}{2\pi}\int_{-\infty}^{\infty}\langle S|\vartheta+\tau/2\rangle\langle\vartheta-\tau/2|S\rangle\exp(-it\tau))d\tau, \qquad (3.36)$$

are the autocorrelation and the Wigner functions, respectively. Let us consider now the absolute value squared of the Fourier transform of the signal S(t):

$$|\langle\omega|S\rangle|^2 = \langle S|\omega\rangle\langle\omega|S\rangle = \int_{-\infty}^{\infty}\int_{-\infty}^{\infty}\langle S|t\rangle\langle t|\omega\rangle\langle\omega|t'\rangle\langle t'|S\rangle dt dt'$$
$$= \frac{1}{2\pi}\int_{-\infty}^{\infty}\int_{-\infty}^{\infty}\langle S|t\rangle\langle t'|S\rangle\exp(i\omega(t-t'))dt dt', \qquad (3.37)$$

following the steps which led to eq. (3.34) we will get:

$$|\langle\omega|S\rangle|^2 = \frac{1}{2\pi}\int_{-\infty}^{\infty}W_2(t,\omega)dt, \qquad (3.38)$$

where:

$$W_2(t,\omega) = \frac{1}{2\pi}\int_{-\infty}^{\infty}\langle S|\omega+\tau/2\rangle\langle\omega-\tau/2|S\rangle\exp(-it\tau))d\tau. \qquad (3.39)$$

One can easily prove by using substitutions that:

$$W_1(t,\omega) = W_2(t,\omega), \qquad (3.40)$$

therefore we have one double distribution function W(t,ω), the Wigner distribution function, satisfying

$$W(t,\omega) = \frac{1}{2\pi} \int_{-\infty}^{\infty} \langle S | \omega + \omega'/2 \rangle \langle \omega - \omega'/2 | S \rangle \exp(-it\omega')) d\omega'$$

$$= \frac{1}{2\pi} \int_{-\infty}^{\infty} \langle S | t + \tau/2 \rangle \langle t - \tau/2 | S \rangle \exp(-i\omega\tau)) d\tau.$$

(3.41)

and (from eqs. (3.34) (3.38)):

$$|\langle \omega | S \rangle|^2 = \frac{1}{2\pi} \int_{-\infty}^{\infty} W(t,\omega) dt, \quad |\langle t | S \rangle|^2 = \frac{1}{2\pi} \int_{-\infty}^{\infty} W(t,\omega) d\omega.$$

(3.42)

In ref. (21) one can find a detailed exposition of the Wigner function and other double distribution functions.

### 4. Non-orthogonal states in subspaces

Let us note that by going to subspaces of time or frequency the orthonormality condition might not be satisfied any more. If we restrict ourselves to the windowed subspace, the rectangular window for example, we obtain:

$$\langle \omega, T | \omega', T \rangle = \int_{-T}^{T} \langle \omega | t \rangle \langle t | \omega' \rangle dt = \frac{1}{2\pi} \int_{-T}^{T} e^{i(\omega'-\omega)t} dt =$$

$$= \frac{e^{i(\omega'-\omega)T} - e^{-i(\omega'-\omega)T}}{2\pi i(\omega-\omega')} = \frac{\sin((\omega-\omega')T)}{\pi(\omega-\omega')},$$

(4.1)

where $|\omega, T\rangle$ are the projected states into the above mentioned subspace, which we will denote by $S_T$. The orthogonality condition is satisfied only if

$$\omega - \omega' = \pm n\pi/T, \quad n = 1, 2, \ldots$$

(4.2)

In order to write eq. (4.1) more explicitly one needs to introduce the projection operator

$$\hat{P}_T = \int_{-T}^{T} |t\rangle \langle t| dt,$$

(4.3)

which projects into the subspace $S_T$. The ket-vector states in this subspace now are:

$$|t, T\rangle = \hat{P}_T |t\rangle = \begin{cases} |t\rangle, & \text{for } |t| \leq T \\ 0, & \text{for } |t| > T, \end{cases}$$

(4.4)

$$|\omega,T\rangle = \hat{P}_T|\omega\rangle = \int_{-T}^{T} |t\rangle\langle t|\omega\rangle dt = \int_{-\infty}^{\infty} |\omega'\rangle \int_{-T}^{T} \langle\omega'|t\rangle\langle t|\omega\rangle dt d\omega'$$
$$= \frac{1}{2\pi}\int_{-\infty}^{\infty} |\omega'\rangle \int_{-T}^{T} e^{it(\omega-\omega')} dt d\omega' = \int_{-\infty}^{\infty} d\omega' \frac{\sin[T(\omega-\omega')]}{\pi(\omega-\omega')}|\omega'\rangle. \quad (4.5)$$

Therefore eq. (4.1) should be written as follows:

$$\langle\omega,T|\omega',T\rangle = \langle\omega|\hat{P}_T\hat{P}_T|\omega'\rangle = \int_{-T}^{T} \langle\omega|t\rangle\langle t|\omega'\rangle dt = \frac{\sin((\omega-\omega')T)}{\pi(\omega-\omega')}, \quad (4.6)$$

The solutions of eq. (4.2), i.e. states which are orthogonal, are:

$$\omega_{n,C} = n\pi/T + C, \quad n = 0,\pm1,\pm2,..., \quad \langle\omega_{n,C},T|\omega_{m,C},T\rangle = \delta_{nm}T/\pi, \quad (4.7)$$

where C is an overall constant and the normalization comes from eq. (4.1). The states of eq. (4.7) form a complete orthogonal set, because we know that the functions

$$\langle t|\omega_{n,0},T\rangle = e^{i\omega_{n,0}t}/\sqrt{2\pi} \quad (4.8)$$

form a complete orthogonal basis for Fourier series in the interval -T≤t≤T. Thus the ket-vector states $|\omega,T\rangle$ are an overcomplete basis in $S_T$. As we shall see later, the relation between the overcomplete basis $|\omega,T\rangle$ and the complete basis of eq. (4.7) is the cornerstone of the sampling theorem. Let us first note that the projection operator $\hat{P}_T$ is the identity operator in the subspace $S_T$:

$$\hat{P}_T|\omega,T\rangle = \hat{P}_T\hat{P}_T|\omega\rangle = \hat{P}_T|\omega\rangle = |\omega,T\rangle. \quad (4.9)$$

Therefore we should expect the completeness relation, taking into account (4.7), to be:

$$\hat{P}_T = \frac{\pi}{T}\sum_{n=-\infty}^{\infty} |\omega_{n,C},T\rangle\langle\omega_{n,C},T|. \quad (4.10)$$

Employing eqs. (4.6) and (4.7), the expansion of the overcomplete basis in terms of the complete one can be obtained in the following way:

$$|\omega,T\rangle = \hat{P}_T|\omega,T\rangle = \frac{\pi}{T}\sum_{n=-\infty}^{\infty} |\omega_{n,C},T\rangle\langle\omega_{n,C},T|\omega,T\rangle$$
$$= \frac{\pi}{T}\sum_{n=-\infty}^{\infty} |\omega_{n,C},T\rangle\frac{\sin[T(\omega_{n,C}-\omega)]}{\pi(\omega_{n,C}-\omega)} = \sum_{n=-\infty}^{\infty} |\omega_{n,C},T\rangle\frac{\sin[T(\omega_{n,C}-\omega)]}{T(\omega_{n,C}-\omega)}, \quad (4.11)$$

from which the following sampling theorem, for a signal f, can be deduced:

$$\langle \omega,T|f \rangle = \sum_{n=-\infty}^{\infty} \frac{\sin[T(\omega_{n,C} - \omega)]}{T(\omega_{n,C} - \omega)} \langle \omega_{n,C},T|f \rangle, \tag{4.12}$$

this result was obtained under the assumption that the signal exists only for times t:

$$|t| \leq T. \tag{4.13}$$

Before closing this section we will prove that the states $|\omega,T\rangle$ form a basis which is overcomplete. We start with the completeness relation (3.8) and modify it according to:

$$\hat{P}_T = \hat{P}_T \hat{I} \hat{P}_T = \int_{-\infty}^{\infty} \hat{P}_T |\omega\rangle\langle\omega|\hat{P}_T d\omega = \int_{-\infty}^{\infty} |\omega,T\rangle\langle\omega,T| d\omega. \tag{4.14}$$

Relation (4.14) is the proof that the states $|\omega,T\rangle$ form a basis in the subspace $S_T$. Indeed any function with the support in $S_T$ have a unique expansion:

$$u(t) = \langle t|u\rangle = \langle t|\hat{P}_T|u\rangle = \int_{-\infty}^{\infty} \langle t|\omega,T\rangle\langle\omega,T|u\rangle d\omega. \tag{4.15}$$

Next we multiply eq. (4.5) from both sides with $\hat{P}_T$ and obtain:

$$\hat{P}_T|\omega,T\rangle = |\omega,T\rangle = \int_{-\infty}^{\infty} d\omega' \frac{\sin[T(\omega - \omega')]}{\pi(\omega - \omega')} |\omega',T\rangle. \tag{4.16}$$

Eq. (4.16) shows that the states $|\omega,T\rangle$ are dependent between themselves, therefore the basis is overcomplete and according to ref. (14) these states satisfy the conditions for being coherent states.

## 5. Sampling theorem for band limited signals

In section 4 the sampling theorem was derived for discrete frequencies for bounded time. In this section we derive the sampling theorem for time states in the case when the frequencies are restricted by the condition

$$|\omega| \leq \omega_B. \tag{5.1}$$

Condition (5.1) restricts the Dirac space to a subspace which we denote as $S_B$. This subspace is reached with the projection operator:

$$\hat{P}_B = \int_{-\omega_B}^{\omega_B} |\omega\rangle\langle\omega| d\omega. \tag{5.2}$$

The ket-vector states in this subspace now are:

$$|\omega, B\rangle = \hat{P}_B |\omega\rangle = \begin{cases} |\omega\rangle, & \text{for } |\omega| \leq \omega_B \\ 0, & \text{for } |\omega| > \omega_B, \end{cases} \qquad (5.3)$$

$$|t, B\rangle = \hat{P}_B |t\rangle = \int_{-B}^{B} |\omega\rangle\langle\omega|t\rangle d\omega = \int_{-\infty}^{\infty} |t'\rangle \int_{-B}^{B} \langle t'|\omega\rangle\langle\omega|t\rangle d\omega dt'$$

$$= \frac{1}{2\pi} \int_{-\infty}^{\infty} |t'\rangle \int_{-B}^{B} e^{i\omega(t'-t)} dt' d\omega = \int_{-\infty}^{\infty} dt' \frac{\sin[B(t'-t)]}{\pi(t'-t)} |t'\rangle. \qquad (5.4)$$

Let us check the scalar products

$$\langle t, B | t', B\rangle = \int_{-\omega_B}^{\omega_B} \langle t|\omega\rangle\langle\omega|t'\rangle d\omega = \frac{\sin((t-t')\omega_B)}{\pi(t-t')}, \qquad (5.5)$$

the orthogonality condition is satisfied only if

$$t - t' = \pm n\pi / \omega_B, \qquad n = 1, 2, \ldots \qquad (5.6)$$

The solutions of eq. (5.6) are the times:

$$t_{n,D} = n\pi / \omega_B + D, \qquad \pm n = 0, 1, 2, \ldots, \qquad (5.8)$$

where D is an overall constant. The corresponding states satisfy the orthogonality condition:

$$\langle t_{n,D}, B | t_{m,D}, B\rangle = \delta_{nm} \omega_B / \pi, \qquad (5.9)$$

and the identity operator for the complete set in $S_B$ is the projection operator:

$$\hat{P}_B = (\pi / \omega_B) \sum_{n=-\infty}^{\infty} |t_{n,D}, B\rangle\langle t_{n,D}, B|. \qquad (5.10)$$

The above results (eqs. 5.5-5.10) allow us to obtain

$$u(t) \equiv \langle t | u\rangle = \langle t | \hat{P}_B | u\rangle = \frac{\pi}{\omega_B} \sum_{n=-\infty}^{\infty} \langle t | t_{n,D}, B\rangle\langle t_{n,D}, B | u\rangle$$

$$= \sum_{n=-\infty}^{\infty} \frac{\sin[(t - t_{n,D})\omega_B]}{(t - t_{n,D})\omega_B} u(t_{n,D}), \qquad (5.11)$$

which for D=0 is the well known sampling theorem, and can be extended for D≠0. Eq. (5.11) can be recast into a more general form:

$$\langle t, B| = \langle t|\hat{P}_B = \frac{\pi}{\omega_B} \sum_{n=-\infty}^{\infty} \langle t, B | t_{n,D}, B\rangle\langle t_{n,D}, B| = \sum_{n=-\infty}^{\infty} \frac{\sin[(t-t_{n,D})\omega_B]}{(t-t_{n,D})\omega_B} \langle t_{n,D}, B|, \qquad (5.11a)$$

or:

$$|t,B\rangle = \hat{P}_B|t\rangle = \frac{\pi}{\omega_B}\sum_{n=-\infty}^{\infty}|t_{n,D},B\rangle\langle t_{n,D},B|t,B\rangle = \sum_{n=-\infty}^{\infty}\frac{\sin[(t-t_{n,D})\omega_B]}{(t-t_{n,D})\omega_B}|t_{n,D},B\rangle. \quad (5.11b)$$

The inverse relation can be obtained in a similar way:

$$|t_{n,D},B\rangle = \hat{P}_B|t_{n,D},B\rangle = \int_{-\infty}^{\infty}|t,B\rangle\langle t,B|t_{n,D},B\rangle dt = \int_{-\infty}^{\infty}dt\frac{\sin[(t-t_{n,D})\omega_B]}{\pi(t-t_{n,D})}|t,B\rangle. \quad (5.11C)$$

Let us decompose the signal u(t) into its odd and even parts:

$$u_+(t) = \tfrac{1}{2}[u(t)+u(-t)], \quad u_-(t) = \tfrac{1}{2}[u(t)-u(-t)], \quad (5.12)$$

and let us denote

$$t_n \equiv t_{n,0}, \quad (5.13)$$

than, from eq. (5.11) the odd and even parts of the signal u(t) can be presented in the following way:

$$\begin{aligned}u_-(t) &= \sum_{n=1}^{\infty}\frac{\sin[(t-t_n)\omega_B]}{(t-t_n)\omega_B}u_-(t_n) - \sum_{n=1}^{\infty}\frac{\sin[(t+t_n)\omega_B]}{(t+t_n)\omega_B}u_-(t_n)\\ u_+(t) &= \frac{\sin(t\omega_B)}{t\omega_B}u_+(0) + \sum_{n=1}^{\infty}\frac{\sin[(t-t_n)\omega_B]}{(t-t_n)\omega_B}u_+(t_n) + \sum_{n=1}^{\infty}\frac{\sin[(t+t_n)\omega_B]}{(t+t_n)\omega_B}u_+(t_n)\end{aligned} \quad (5.14)$$

In eq. (5.14) the signal is sampled only for non-negative times. This is an interesting possibility if the time parity is known and fixed.

We can consider the sampling theorem from a different point of view if we substitute eqs. (5.8) and eq. (5.13) into eq. (5.11), we obtain:

$$u(t) = \sum_{n=-\infty}^{\infty}\frac{(-1)^n \sin(\omega_B t)}{\omega_B t - n\pi}u(t_n), \quad (5.15)$$

or:

$$\frac{u(t)}{\sin(\omega_B t)} = \sum_{n=-\infty}^{\infty}\frac{(-1)^n}{\omega_B t - n\pi}u(t_n). \quad (5.16)$$

Thus eq. (5.16) is the pole expansion of its left hand side, and it is valid only if u(t) does not have singularities in t except at infinity. It has also to comply to the Mittag-Leffler 's theorem [26],

We can still view the sampling theorem from other angle. In eqs. (5.6)-(5.10) we have different orthogonal bases characterized by the parameter D, let us call it B(D). Every point t can be reached by t=$t_n$+D for some values of n and D. Therefore we can present the sampling theorem from that point of view by expanding the base vectors of B(D) in terms of the base vectors of B(D=0). For this we need only to use eqs (5.11). We obtain:

$$\langle t_{n,D} | t_m \rangle = \frac{\sin[(t_{n,D} - t_m)\omega_B]}{\pi(t_{n,D} - t_m)}, \quad (5.17)$$

and the sampling theorem takes the form:

$$t = t_{k,D}$$

$$\langle t | u \rangle = \langle t_{k,D} | u \rangle = \sum_{n=-\infty}^{\infty} \frac{\sin[(t_{k,D} - t_n)\omega_B]}{(t_{k,D} - t_n)\omega_B} u(t_n), \quad (5.18)$$

where now only discrete orthogonal sets are involved. Eq. (5.18) a consequence of isometric isomorphism between the bases labeled (k,D) and those labeled (k,0).

It is interesting to note that all band limited functions, $f(t) \equiv \langle t | P_B | f \rangle$, have to satisfy the identity given by eqs. (A.10) and (A.11) of Appendix A. Using eq. (5.5), we obtain

$$\begin{aligned}\langle t | P_B | f \rangle &= \int_{-\infty}^{\infty} \langle t | P_B | t' \rangle \langle t' | P_B | f \rangle dt' \\ &= \int_{-\infty}^{\infty} \frac{\sin(\omega_B(t-t'))}{\pi(t-t')} \langle t' | P_B | f \rangle dt'.\end{aligned} \quad (5.19)$$

The identity (5.19) was employed in refs. [27] and [28] in order to derive an expansion for band limited functions in a time window.

## 6. Some consequences of the sampling theorem and some ambiguities

First let us note, that in order to avoid some ambiguities, the condition (5.1) has to be changed to

$$|\omega| < \omega_B, \quad (6.1)$$

otherwise we can add to the eq. (5.11) any function which is zero at the interpolation points $t_{n,D}$. Thus we can get an ambiguity of up to a function having zeros at $t = t_{n,D}$. Such a function will have a periodicity of $\omega = \omega_B$. The condition (6.1) will eliminate that possibility

A consequence of eq. (5.10) is the replacement of the Fourier transform eq. (3.3) by the sum

$$\langle\omega|f\rangle = \frac{\pi}{\omega_B} \sum_{n=-\infty}^{\infty} \langle\omega|t_{n,D}\rangle\langle t_{n,D}|f\rangle = \frac{\sqrt{\pi}}{\sqrt{2}\omega_B} \sum_{n=-\infty}^{\infty} e^{-i\omega t_{n,D}} f(t_{n,D}), \tag{6.2}$$

where we continue to use the normalization given by eq. (3.6)

$$\langle\omega|t_{n,D}\rangle = \frac{1}{\sqrt{2\pi}} e^{i\omega t_{n,D}}. \tag{6.3}$$

The sampling theorem is recovered from eq. (6.2) by the integration:

$$\langle t|f\rangle = \int_{-\omega_B}^{\omega_B} \langle t|\omega\rangle\langle\omega|f\rangle d\omega. \tag{6.4}$$

We shall derive, using eqs. (5.10) and (5.8), an important relation to be used later on

$$\delta(\omega-\omega') = \langle\omega|\omega'\rangle = \frac{\pi}{\omega_B} \sum_{n=-\infty}^{\infty} \langle\omega|t_{n,0}\rangle\langle t_{n,0}|\omega'\rangle = \frac{1}{2\omega_B} \sum_{n=-\infty}^{\infty} e^{in\pi(\omega'-\omega)/\omega_B}. \tag{6.5}$$

The right hand side of eq. (6.5) is a periodic function of period $2\omega_B$, therefore we infer that it should repeat itself every period, thus we should have:

$$\frac{1}{2\omega_B} \sum_{n=-\infty}^{\infty} e^{in\pi\omega/\omega_B} = \sum_{m=-\infty}^{\infty} \delta(\omega - 2\omega_B m). \tag{6.6}$$

The last relation is proved more rigorously in ref. [3].

Caution has to be taken in using eq. (6.2) which has a periodicity of $\omega = 2\omega_B$, in contradiction to eqs. (5.1) and (6.1). The restricted limits integration (6.4) bring us back to eqs. (5.1) and (6.1). Therefore eq. (6.2) can be used without ambiguities only under the condition (6.1).

Let us consider an example:

$$F(t) \equiv \langle t|F\rangle = e^{i\omega_0 t}, \quad \omega_0 > 0, \text{ (real)}, \tag{6.7}$$

and let us check eq. (5.11) by first evaluating eq. (6.2) (with D=0), than eq. (6.4).

$$\langle\omega|F\rangle = \frac{\sqrt{\pi}}{\sqrt{2}\omega_B} \sum_{n=-\infty}^{\infty} e^{i(\omega_0-\omega)t_n} = \sqrt{2\pi} \sum_{m=-\infty}^{\infty} \delta(\omega_0 - \omega - 2m\omega_B), \quad |\omega_0| < \omega_B. \tag{6.8}$$

where eq. (6.6) was used. Insertion of eq. (6.8) into eq. (6.4) recovers the correct result i.e. eq. (6.7). If instead of condition (6.1) the condition (5.1) will be used, than one is allowed to take $\omega_B = \omega_0$ and instead of eq. (6.8) we obtain:

$$\langle\omega|F\rangle = \frac{\sqrt{\pi}}{\sqrt{2}\omega_B} \sum_{n=-\infty}^{\infty} e^{i(\omega_0-\omega)t_n} = \sqrt{2\pi} \sum_{m=-\infty}^{\infty} \delta(\omega_B - \omega - 2m\omega_B), \quad \omega_B = \omega_0, \qquad (6.9)$$

and after substituting in eq. (6.4) we obtain (the delta function is an even function):

$$\langle t|F\rangle = \int_{-\omega_B}^{\omega_B} e^{i\omega t}[\delta(\omega-\omega_B) + \delta(\omega+\omega_B)]d\omega \neq e^{i\omega_0 t}, \qquad (6.10)$$

i.e. we do not recover the result (6.7).

Now we shall derive another application of eq. (6.6). Substituting the Fourier transform (3.2) into eq. (6.2) we obtain:

$$\frac{\sqrt{\pi}}{\sqrt{2}\omega_B} \sum_{n=-\infty}^{\infty} e^{-i\omega t_{n,0}} f(t_{n,0}) = \frac{1}{2\omega_B} \int_{-\infty}^{\infty} \left( \sum_{n=-\infty}^{\infty} e^{i(\omega'-\omega)t_{n,0}} \right) F(\omega')d\omega'$$
$$= \sum_{n=-\infty}^{\infty} F(\omega - 2\omega_B n), \qquad (6.11)$$

which coincides with eq. (6.2) under the constriction (6.1). If no constraints are imposed on the frequencies, eq. (6.11) and its complex conjugate are the manifestation of the aliasing phenomenon.

The main ingredients leading to the sampling theorem are eqs. (5.5), (5.9) and (5.10). A consequence of eq. (5.9), using eq. (5.5), is

$$\langle t_{n,D}|t_{m,D}\rangle = \int_{-\infty}^{\infty} \langle t_{n,D}|t\rangle\langle t|t_{m,D}\rangle dt = \delta_{nm}\omega_B/\pi. \qquad (6.12)$$

or

$$\int_{-\infty}^{\infty} \frac{\sin[(t_{n,D}-t)\omega_B]}{(t_{n,D}-t)\omega_B} \frac{\sin[(t_{m,D}-t)\omega_B]}{(t_{m,D}-t)\omega_B} dt = \frac{\delta_{nm}}{\omega_B \pi}. \qquad (6.13)$$

### 7. Spectral analysis on the half-infinite time axis.

In the sampling theorem and related equations of section 5 the time is running from minus to plus infinity. In practice we start to collect data at a definite time, which can be chosen to be t=0. Therefore half axis time analysis (t>0) is more related to experiment than whole time axis analysis. Moreover we deal with real signals. For real signals the Fourier transform eq. (3.1) will have the following property:

$$F^*(\omega) = F(-\omega). \qquad (7.1)$$

Let us define the even and odd parts of the signal F(t) of eq. (3.2) as:

$$f_+(t) = \tfrac{1}{2}[f(t) + f(-t)], \quad f_-(t) = \tfrac{1}{2}[f(t) - f(-t)], \qquad (7.2)$$

than, using eqs. (3.1), (3.2), (7.1), (7.2) we obtain the sine and cosine Fourier transforms:

$$f_+(t) = \frac{1}{\sqrt{2\pi}} \int_{-\infty}^{+\infty} \cos(\omega t) \tfrac{1}{2}[F(\omega) + F^*(\omega)] d\omega = \sqrt{\frac{2}{\pi}} \int_0^{+\infty} \cos(\omega t) \operatorname{Re}[F(\omega)] d\omega, \qquad (7.3)$$

and in a similar way:

$$f_-(t) = -\sqrt{\frac{2}{\pi}} \int_0^{+\infty} \sin(\omega t) \operatorname{Im}[F(\omega)] d\omega, \qquad (7.4)$$

$$\operatorname{Re}[F(\omega)] = \sqrt{\frac{2}{\pi}} \int_0^{+\infty} \cos(\omega t)[f_+(t)] dt, \qquad (7.5)$$

$$\operatorname{Im}[F(\omega)] = -\sqrt{\frac{2}{\pi}} \int_0^{+\infty} \sin(\omega t)[f_-(t)] dt. \qquad (7.6)$$

In terms of the Dirac formalism the sine and cosine transforms can be formulated as follows: the vector space with the basis $|t\rangle$ is decomposed into two orthogonal subspaces $|t, P = 1\rangle$ (with projection P=1 into the even functions of t) and $|t, P = -1\rangle$ (with projection P=-1 into odd functions of t). Now, because of the symmetry of even and odd functions in t →-t, all the information can be stored in t≥0. Thus any function of t can be decomposed into its even and odd parts, as in eq. (7.2):

$$f(t) \equiv \langle t | f \rangle = [f_+(t) + f_-(t)] \equiv [\langle t, P = 1 | f \rangle + \langle t, P = -1 | f \rangle], \qquad (7.7)$$

from which we deduce that:

$$|t\rangle = |t, P = 1\rangle + |t, P = -1\rangle. \qquad (7.8)$$

Let us note from eqs. (3.6), (3.7), (7.8), and (7.9) that:

$$\begin{aligned}
&\langle t, P = 1 | \omega, P = 1 \rangle = \frac{1}{\sqrt{2\pi}} \cos(\omega t), \quad \langle t, P = -1 | \omega, P = -1 \rangle = \frac{i}{\sqrt{2\pi}} \sin(\omega t), \\
&\langle t, P = 1 | t', = P - 1 \rangle = 0, \\
&\langle t, P = 1 | t', P = 1 \rangle = \langle t, P = -1 | t', P = -1 \rangle = \tfrac{1}{2} \delta(t - t'),
\end{aligned} \qquad (7.9)$$

In a similar way we can define even and odd parity states of frequency.

The identity operators will acquire the forms:

$$\hat{I} = \int_{-\infty}^{\infty} |t, P = 1\rangle\langle t, P = 1| dt + \int_{-\infty}^{\infty} |t, P = -1\rangle\langle t, P = -1| dt$$

$$\hat{I} = \int_{-\infty}^{\infty} |\omega, P = 1\rangle\langle \omega, P = 1| d\omega + \int_{-\infty}^{\infty} |\omega, P = -1\rangle\langle \omega, P = -1| d\omega$$

(7.10)

For real signals, eqs. (7.3)-(7.6) will acquire the following form:

$$\langle t, P = 1 | f \rangle = 2 \int_0^{\infty} \langle t, P = 1 | \omega, P = 1 \rangle \langle \omega, P = 1 | f \rangle d\omega \qquad (7.3a)$$

$$\langle t, P = -1 | f \rangle = 2 \int_0^{\infty} \langle t, P = -1 | \omega, P = -1 \rangle \langle \omega, P = -1 | f \rangle d\omega, \qquad (7.4a)$$

$$\langle \omega, P = 1 | f \rangle = 2 \int_0^{\infty} \langle \omega, P = 1 | \omega, P = 1 \rangle \langle t, P = 1 | f \rangle dt, \qquad (7.5a)$$

$$\langle \omega, P = -1 | f \rangle = 2 \int_0^{\infty} \langle \omega, P = -1 | \omega, P = -1 \rangle \langle t, P = -1 | f \rangle dt \qquad (7.6a)$$

We see in eqs. (7.3)-(7.6), (7.3a)-(7.6a) that in the sine and cosine transforms the integration is for positive time, but the real signals must have a definite parity, i.e. they must be either even (P=1) or odd (P=-1) functions of t. Let us now consider the general case with no definite parity.

As we have pointed out, in an experiment we measure the signal, starting to collect data at a definite time, which can be taken as t=0. Before that time our apparatus is idle. Let us consider for simplicity the two signals:

$$\langle t | F_1 \rangle = \frac{1}{\sqrt{2\pi}} \exp(i\omega_0 t) \equiv \langle t | \omega_0 \rangle,$$

$$\langle t | F_2 \rangle = \begin{cases} 0, & \text{for } t < 0, \\ \frac{1}{\sqrt{2\pi}} \exp(i\omega_0 t), & \text{forýt} \geq 0, \end{cases}$$

(7.11)

and evaluate their Fourier transforms, from eq. (3.3):

$$\langle \omega | F_1 \rangle = \int_{-\infty}^{\infty} \langle \omega | t \rangle \langle t | \omega_0 \rangle dt = \delta(\omega - \omega_0),$$

$$\langle \omega | F_2 \rangle = \frac{1}{2\pi} \int_0^{\infty} \exp[i(\omega_0 - \omega)t] dt = \frac{1}{2\pi i} \frac{1}{\omega - \omega_0} + \tfrac{1}{2}\delta(\omega_0 - \omega),$$

(7.12)

Thus in a positive time spectral analysis a single frequency signal can be represented by a pole in the frequency with a possibility of using rational approximations. Let us introduce the positive time projection operator

$$\hat{t}_+ = \int_0^\infty |t\rangle\langle t|dt, \qquad (7.13)$$

and apply it on the kets and bras of our vector space. We will note that the frequency states will no longer be orthogonal:

$$\langle\omega|\hat{t}_+|\omega'\rangle = \int_0^\infty \langle\omega|t\rangle\langle t|\omega'\rangle dt = \frac{1}{2\pi i}\frac{1}{\omega-\omega'} + \frac{1}{2}\delta(\omega-\omega'). \qquad (7.14)$$

Let us denote

$$A(\omega) \equiv \langle\omega|\hat{t}_+|F\rangle = \int_0^\infty \langle\omega|t\rangle\langle t|\hat{t}_+|F\rangle dt = \frac{1}{\sqrt{2\pi}}\int_0^\infty \exp(-i\omega t)F(t)dt, \qquad (7.15)$$

for the signal $F(t)$. The inverse relation (for positive t) is:

$$F(t) = \langle t|\hat{t}_+|F\rangle = \int_{-\infty}^\infty \langle t|\hat{t}_+|\omega\rangle\langle\omega|\hat{t}_+|F\rangle d\omega = \frac{1}{\sqrt{2\pi}}\int_{-\infty}^\infty \exp(i\omega t)A(\omega)d\omega. \qquad (7.16)$$

From eq. (A.10) of Appendix A and eq. (7.12) we infer that

$$A(\omega) = \frac{1}{\pi i}\int_{-\infty}^\infty \frac{A(\omega')}{\omega'-\omega}d\omega', \qquad (7.17)$$

which is a Hilbert transform for the real and imaginary parts of A:

$$\begin{aligned}\text{Re}[A(\omega)] &= \frac{1}{\pi}\int_{-\infty}^\infty \frac{\text{Im}[A(\omega')]}{\omega'-\omega}d\omega', \\ \text{Im}[A(\omega)] &= \frac{-1}{\pi}\int_{-\infty}^\infty \frac{\text{Re}[A(\omega')]}{\omega'-\omega}d\omega'.\end{aligned} \qquad (7.18)$$

In eqs. (7.15-7.18) fast enough decay as $\omega\to\infty$ is required for the existence of the integrals.

Let us consider the case of the frequency bounded by eq. (6.1) and consider the sampling theorem for positive time signals. As in the case of eq. (6.2) one can replace the transform (7.15) by the sum:

$$A(\omega) \equiv \langle \omega | \hat{t}_+ | F \rangle = \frac{\pi}{\omega_B} \sum_{n=-\infty}^{\infty} \langle \omega | \hat{t}_+ | t_{n,D} \rangle \langle t_{n,D} | F \rangle$$
$$= \frac{\pi}{\omega_B} \sum_{n+D \geq 0} \left[ \int_0^\infty \langle \omega | t \rangle \langle t | t_{n,D} \rangle dt \right] \langle t_{n,D} | F \rangle, \tag{7.19}$$

which is not the same as the (similar to eq. (6.2)) expression:

$$B(\omega) = \frac{\pi}{\omega_B} \sum_{n+D \geq 0}^{\infty} \langle \omega | t_{n,D} \rangle \langle t_{n,D} | F \rangle = \frac{\sqrt{\pi}}{\sqrt{2}\omega_B} \sum_{n+D \geq 0}^{\infty} e^{-i\omega t_{n,D}} F(t_{n,D}). \tag{7.20}$$

Let us consider as an example the second signal of eq. (7.11) for which eq. (7.20), with D=0, becomes:

$$B(\omega) = \frac{1}{2\omega_B} \sum_{n=0}^{\infty} e^{i(\omega_0 - \omega)n\pi/\omega_B} = \frac{1}{2\omega_B} \frac{1}{1 - \exp(i(\omega_0 - \omega)\pi/\omega_B)}, \tag{7.21}$$

which differs from the second expression in eq. (7.12), although both expressions have the same pole and residuum. We mention this because eq. (7.20) is the basis for one sided Z-transform [5-8]. Instead, expression (7.19) should be used. Expression (7.20) is a good approximation to (7.19) in the sense that it has the same poles and residua. In eq. (7.19) there appear new coefficients:

$$C(\omega, t_{n,D}) = \int_0^\infty \langle \omega | t \rangle \langle t | t_{n,D} \rangle dt = \frac{1}{\sqrt{2\pi}} \int_0^\infty e^{-i\omega t} \frac{\sin((t - t_{n,D})\omega_B)}{\pi(t - t_{n,D})} dt, \tag{7.22}$$

this expression can be brought, with the help of eqs. (4.6) and (7.14) to the form:

$$C(\omega, t_{n,D}) = \frac{1}{\sqrt{2\pi}} \int_{-\omega_B}^{\omega_B} \frac{\exp(-i\omega' t_{n,D})}{2\pi i(\omega' - \omega)} d\omega'. \tag{7.23}$$

## 8. Summary and conclusions

We have demonstrated in this paper that the Dirac representation theory can be effectively adjusted and applied to signal theory. The advantages of the Dirac representation theory is its clarity, transparency and generality. It is suitable to be used with generalized functions and general transforms. The use of the identity operator and projection operators simplify greatly the calculations and derivations. Using these operators we introduced in Appendix A, a large class of functions, which were called the incomplete delta functions, as they have similar properties to the Dirac delta function and converge to it in a limiting process. They have been used throughout the paper in order to simplify derivations. The case of wavelets deviates somewhat from the standard procedure, and is collected in Appendix C.

The scalar products (brackets) of Dirac are written in a way which does not depend on the choice of the coordinates. In the process of representing the signal theory with the brackets we found interesting features of scalar products which were related to the change of validity domains of the conjugate coordinates (in signal theory they are the time and frequency, in quantum mechanics the coordinates and momenta). An orthogonal basis which spanned the time space ceased to be orthogonal after constraining the domain of frequencies (bandwidth) and becomes an overcomplete. But in this case there exist a sub-bases of discrete times which are orthogonal and complete. The completeness of the two bases, one for all frequencies (the overcomplete non-orthogonal basis), the other in the bandwidth, eq. (4.10), allows the derivation of the sampling theorem relating signals in the discrete time (complete) basis to signals for all times (with the overcomplete basis - eq. (4.11)). Thus the relation between the overcomplete bases and a complete one is the essence of the sampling theorem. Moreover eq. (4.10) allows us to replace the continuous time integration by a discrete summation in all time integrations, this we could do in the Fourier transform, eq. (6.2), in the wavelets transform (Appendix C) and in the case of signals existing for positive time - eq. (7.19) .

The idea of viewing the sampling theorem as the relation between an overcomplete basis and a complete one can be extended to quantum mechanics. We are in the process of completing a work in this direction.

## Appendix A
**The incomplete delta functions and the delta function of Dirac**

Let us consider two complete continuous bases:
$$\langle x|x'\rangle = \delta(x-x'), \quad \langle y|y'\rangle = \delta(y-y'), \quad -\infty \leq y \leq \infty; \; -\infty \leq x \leq \infty, \tag{A.1}$$
linked together by the Fourier transform (eq. (3.6)) :

$$\langle x | y \rangle = \frac{1}{\sqrt{2\pi}} \exp(ixy). \tag{A.2}$$

where $\delta(x)$ is the delta function of Dirac. The completeness relation is given by

$$I = \int_{-\infty}^{\infty} |x\rangle\langle x| dx = \int_{-\infty}^{\infty} |y\rangle\langle y| dy, \tag{A.3}$$

where I is the identity operator. Let us consider the following set of functions:

$$\delta(x - x', a) = \int_{-a}^{a} \langle x | y \rangle \langle y | x' \rangle dy$$

$$= \frac{1}{2\pi} \int_{-a}^{a} e^{ixy} dy = \frac{a}{\pi} \frac{\sin(ax)}{ax}, \quad a > 0 \tag{A.4}$$

from which the Dirac delta function $\delta(x)$ can be obtained in the limiting process:

$$\delta(x) = \lim_{a \to \infty} \delta(x, a). \tag{A.5}$$

The Dirac delta function (distribution, or generalized function) has the following properties:

$$\delta(x) \equiv \lim_{a \to \infty} \delta_a(x) = \begin{cases} 0, & \text{for } x \neq 0, \\ \infty, & \text{for } x = 0. \end{cases}$$

$$\int_{-\infty}^{\infty} \delta(x) dx \equiv \lim_{a \to \infty} \int_{-\infty}^{\infty} \delta_a(x) dx = 1, \tag{A.6}$$

$$\int_{-\infty}^{\infty} f(x) \delta(x - x_0) dx \equiv \lim_{a \to \infty} \int_{-\infty}^{\infty} f(x) \delta_a(x - x_0) dx = f(x_0).$$

One can generalize the above procedure to a larger class of functions which we will call "incomplete delta functions". Let P be a projection operator, such that by changing its parameters, in the limiting process, it converges to the identity operator. An example is eq. (A.4), in which the following projection operator was used:

$$P(a) = \int_{-a}^{a} |y\rangle\langle y| dy; \quad \lim_{a \to \infty} P(a) = I. \tag{A.7}$$

Let us define the incomplete delta function as

$$\delta(x - x', P) = (\langle x|P)(P|x'\rangle), \quad \text{such that} \quad \delta(x - x', P \to I) = \delta(x - x'), \tag{A.8}$$

i.e. the states denoted by x are projected to a subspace with the projection operator P. The projection operator satisfies (see eqs. (2.11a) and (2.23)): $P=P^2$ and $P=P^\dagger$, which can be employed in order to derive general properties of the incomplete delta functions. We can prove that similar equations to eq. (A.6) are satisfied for the incomplete delta functions. Inserting the identity operator of eq. (A.3) into eq. (A.8) we obtain:

$$\delta(x-x',P) = (\langle x|P\rangle)\, I\, (P|x'\rangle) = \int_{-\infty}^{\infty} \langle x|PP|x''\rangle\langle x''|PP|x''\rangle dx''$$
$$= \int_{-\infty}^{\infty} \delta(x-x'',P)\delta(x''-x',P)dx''. \tag{A.9}$$

Let f(x) be a function with a support in the projected subspace, than:

$$f(x) = \langle x|f\rangle = \langle x|P|f\rangle = \langle x|PIP|f\rangle = \int_{-\infty}^{\infty} \langle x|PP|x'\rangle\langle x'|P|f\rangle dx'$$
$$= \int_{-\infty}^{\infty} \delta(x-x',P)f(x')dx'. \tag{A.10}$$

For non restricted functions still the following identity holds:

$$\langle x|P|f\rangle = \int_{-\infty}^{\infty} \delta(x-x',P)\langle x'|P|f\rangle dx'. \tag{A.11}$$

The identities (A.10) may have applications in testing hypotheses that a given function has a support in a restricted subspace.

One can note from eq. (A.4) that:

$$\delta(0,a) = a/\pi. \tag{A.12}$$

We shall prove the important property of the incomplete delta function (A.4):

$$\delta(bx,a) = \frac{1}{|b|}\delta(x,a|b|), \tag{A.13}$$

where b is a real number. For b>0 we have:

$$\delta(bx,a) = \frac{1}{2\pi}\int_{-a}^{a} e^{ixby}dy = \frac{1}{2\pi b}\int_{-ab}^{ab} e^{ixz}dz = \frac{1}{b}\delta(x,ab). \tag{A.14}$$

For b<0 we obtain:

$$\delta(bx,a) = \frac{1}{2\pi}\int_{-a}^{a} e^{ixby}dy = \frac{1}{2\pi b}\int_{-ab}^{ab} e^{ixz}dz = \frac{-1}{2\pi|b|}\int_{a|b|}^{-a|b|} e^{ixz}dz = \frac{1}{|b|}\delta(x,a|b|),$$

which, together with eq. (A.14), proves eq. (A.13). From eq. (A.13) we can get:

$$\delta(bx) = \frac{1}{|b|}\delta(x) . \tag{A.15}$$

Eq. (A.15) can be generalized to include functions f(x) having zeros at $x=x_n$:

$$\delta(f(x)) = \sum_n \frac{1}{|f'(x_n)|}\delta(x-x_n) , \tag{A.16}$$

where near the zeros:

$$f(x) \approx f'(x_n)(x-x_n) . \tag{A.17}$$

## Appendix B.

**Analysis of respirator's cycle.**

The working cycle of period P of the respirator in the operation room of the Gynecology Department, Soroka Medical Center, Beer-Sheva, Israel, was approximately as follows:

$$V(t) = \begin{cases} A\sin(2\pi\varphi_1 t) & \text{for} \quad 0 < t < t_1 \quad \text{(pumping)}, \\ 0, & \text{for} \quad t_1 < t < P \quad \text{(idle)}, \end{cases} \tag{B.1}$$

where V was the volume of the pumped air, A is a constant volume, $t_1 = 3\,\text{sec}$, $\varphi_1 = (1/6)\,\text{Hz}$ and the period of P was changing with the anesthetic procedure. Most of the time it was P=5sec, within the range 3.3sec<P<16sec.

For practical calculation purposes it will be useful to translate $t \to t + P/2$ by half a period to obtain V(t) in a symmetric form for -P/2 < t < P/2 :

$$V(t) = \begin{cases} A\sin(2\pi\varphi_1|t-t_2|) & \text{for} \quad t_2 < |t| \quad \text{(pumping)}, \\ 0, & \text{for} \quad -t_2 < t < t_2 \quad \text{(idle)}, \end{cases} \tag{B.2}$$

where

$$t_2 = (P-t_1)/2 .$$

The above function is periodic, with a period P, and is an even function of t. Therefore a more restricted basis than (3.6) may be more suitable, namely the basis of even functions with period P:

$$\langle t | n, P \rangle = \cos(2n\pi t/P), \quad n=0,1,2,\ldots , \tag{B.3}$$

and $V(t) \equiv \langle t | V \rangle$ can be expanded in a following Fourier expansion:

$$V(t) = \langle t | V \rangle = \sum_{n=0}^{\infty} \langle t | n, P \rangle \langle n, P | V \rangle$$
$$= \frac{a_0}{2} + \sum_{n=1}^{n=\infty} a_n \cos(2n\pi t / P),$$
(B.4)

where $\langle t | n, P \rangle$ are periodic symmetric functions of t, orthonormal in the interval [-P/2, P/2] and

$$a_n = \frac{1}{P} \int_{t_2}^{P/2} V(t) \cos(2n\pi t / P) dt$$
$$= \frac{2A}{\pi P} \left\{ \frac{\cos 2\pi[(f_n - \varphi_1)t + \varphi_1 t_2]}{f_n - \varphi_1} - \frac{\cos 2\pi[(f_n + \varphi_1)t - \varphi_1 t_2]}{f_n + \varphi_1} \right\}_{t=t_2}^{t=P/2},$$
(B.5)

and

$$f_n = n / P.$$
(B.6)

Here $f_1 = 1/P$ is the basic frequency and $f_n$ (n>1) are the higher harmonics. In practice measurements are taken for finite intervals. In this case the spectral analysis should be modified according to the procedure outlined in the main text eqs. (3.9) and (3.10). With the finite interval of duration 2T the spectral analysis of eq. (B.3) (with F(t)=V(t) in eq. (3.9) and remembering that $\omega = 2\pi f$) becomes:

$$G(\omega)_T = \frac{a_0 \sin(\omega T)}{\omega} + \sum_{n=1}^{\infty} a_n \left\{ \frac{\sin[(\omega - \omega_n)T]}{\omega - \omega_n} + \frac{\sin[(\omega + \omega_n)T]}{\omega + \omega_n} \right\},$$
(B.7)

where $\omega_n = 2\pi f_n = 2\pi n / P$.

## Appendix C
**Dirac Formalism as a framework for wavelets**

We can show that the Dirac formalism can be applied to the wavelets transforms [29] of the signal f(t), which are defined in the following way:

$$(W_\psi f)(b, a) = |a|^{-\frac{1}{2}} \int_{-\infty}^{\infty} f(t) \psi^* \left( \frac{t-b}{a} \right) dt.$$
(C.1)

Above in eq. (C.1) $\psi$ is a square integrable function, the "basic wavelet". In order to apply the Dirac formalism we have to assume that the function $\psi$ is generated by an operator W which is represented in eq. (C.1) by its matrix elements:

$$\langle t|W|b,a\rangle = |a|^{-\frac{1}{2}} \psi\left(\frac{t-b}{a}\right). \tag{C.2}$$

Than eq. (C.1), the wavelet transform of f(t), can be represented as:

$$\langle b,a|W|f\rangle = \int_{-\infty}^{\infty} \langle b,a|W|t\rangle\langle t|f\rangle dt. \tag{C.3}$$

The discrete wavelet transform of Daubechies [30] can be treated in a similar way. Let us denote by $W_D$ the operator generating the wavelets, with matrix elements:

$$\langle t|W_D|j,k\rangle \equiv \psi_{j,k}(t) = 2^{j/2}\psi(2^j t - k). \tag{C.4}$$

They form an orthonormal basis:

$$\int_{-\infty}^{\infty} \langle l,m|W_D|t\rangle\langle t|W_D|j,k\rangle dt = \delta_{j,l}\cdot\delta_{k,m}, \tag{C.5}$$

with the identity operator:

$$I = \sum_{j,k=-\infty}^{\infty} W_D|j,k\rangle\langle j,k|W_D. \tag{C.6}$$

The signal f(t) can be expanded using eq. (C.6)

$$f(t) \equiv \langle t|f\rangle = \sum_{j,k=-\infty}^{\infty} \langle t|W_D|j,k\rangle\langle j,k|W_D|f\rangle, \tag{C.7}$$

where :

$$\langle j,k|W_D|f\rangle = \int_{-\infty}^{\infty} \langle j,k|W_D|t\rangle\langle t|f\rangle dt. \tag{C.8}$$

Other discrete wavelet transforms [31] can go along an identical path. We shall now look for operators which can generate the wavelet transformation of equation (C.2). Let us assume that the operator V(a,b) has the following property:

$$V(a,b)f(t) = \frac{1}{\sqrt{a}} f\left(\frac{t-b}{a}\right). \tag{C.9}$$

Such an operator can be formed from the following operators [21]:

$$V(a,b) = e^{-i(\ln a)C} e^{-ib\Phi}, \tag{C.10}$$

where C is the scale operator and $\Phi$ the frequency operator defined by:

$$C = \frac{1}{2i}\left(t\frac{d}{dt} + \frac{d}{dt}t\right), \qquad \Phi = \frac{1}{i}\frac{d}{dt}. \tag{C.11}$$

The result (C.9) is obtained by translation:

$$e^{-ib\Phi}f(t) = f(t-b), \tag{C.12}$$

and by compression [21]:

$$e^{-i(\ln a)C}f(t) = \frac{1}{\sqrt{a}}f\left(\frac{t}{a}\right). \tag{C.13}$$

The transformation V of eq. (C.10) is unitary and thus the normalization is preserved under its action. Using results of ref. (10) we can represent the completeness relation and the inverse transform in the following way. The completeness relation can be represented by the identity operator:

$$I = \frac{1}{C_\psi}\iint \Psi|b,a\rangle\langle b,a|\Psi \frac{db\, da}{a^2}, \tag{C.14}$$

where the normalization constant $C_\psi$ is defined by:

$$C_\psi = \int_{-\infty}^{\infty} \frac{|\hat{\psi}(\omega)|^2}{|\omega|}d\omega < \infty, \tag{C.15}$$

and $\hat{\psi}(\omega)$ is the Fourier transform of $\psi(t)$ of eq. (C.2). Using eq. (C.14) we can get the inverse transform of eq. (C.3):

$$f(t) \equiv \langle t|f\rangle = \frac{1}{C_\psi}\iint \langle t|\Psi|b,a\rangle\langle b,a|\Psi|f\rangle \frac{db\, da}{a^2}. \tag{C.16}$$

In a similar way the wavelet transform (2.19) can be modified to:

$$\langle b,a|\Psi|f\rangle = (\pi/\omega_B)\sum_{n=-\infty}^{\infty}\langle b,a|\Psi|t_{n,D}\rangle\langle t_{n,D}|f\rangle, \tag{5.3a}$$

and the wavelet transform (2.24) to:

$$\langle j,k|Da|f\rangle = (\pi/\omega_B)\sum_{n=-\infty}^{\infty}\langle j,k|Da|t_{n,D}\rangle\langle t_{n,D}|f\rangle. \tag{5.3b}$$